# Title: Chiral Balls: Knotted Structures with Both Chirality and Three-dimensional Rotational Symmetry

**Authors:** Wending Mai[1, 2*], Chunxu Mao[2], Lei Kang[2],

Douglas H. Werner[2], Jun Hu[1], and Yifan Chen[1]

**Affiliations:**

[1] School of Life Science and Technology, The University of Electronic Science and Technology of China, Chengdu, 611731 China

[2] Electrical Engineering Department, The Pennsylvania State University, University Park, PA 16802 USA

[3] School of Electronic Science and Engineering, The University of Electronic Science and Technology of China, Chengdu, 611731 China

Correspondence to: Wending Mai wdm@ieee.org

**Abstract**: Knots have been put forward to explain various physical phenomena because of their topological stability. Nevertheless, very few works have reported on the exotic symmetry properties that certain knots possess. Here we reveal an exceptional form of symmetry for a family of knots that are both chiral and three-dimensional (3-D) rotationally symmetric about every axis of a standard Cartesian coordinate system. We call these unique knotted structures *chiral balls*. Our work shows that, besides topological stability, certain knots possess this exotic form of symmetry, which has not been previously reported. Because of their unique properties, chiral balls are expected to not only have a profound impact on the fields of mathematics and physics but also potentially far beyond.

**Main Text:** Knots have played an important role throughout human history, for nets, traps, clothing, as well as a means of recording and communicating information. Mathematically, a knot is defined as a closed curve in three-dimensional space that does not intersect itself anywhere (*1*). Any deformations of the closed knotted curve through space that do not allow the curve to pass through itself are considered, from a topological point of view, to be the same knot. In an early scientific application, knots were considered as one possible explanation for the elementary particles in Kelvin's atomic theory (*2*). More recently, knots have found a much wider range of interesting and useful applications in the fields of chemistry, physics, biology, and engineering. Examples of such diverse applications include fluid dynamics (*3*), photonics (*4*), molecular structure of DNA and other complex molecules (*5–7*), electrostatic and magnetostatic fields (*8*, *9*), and electromagnetic scattering properties (*10-16*).

However, few works have been reported that exploit the exotic symmetry of knots. Symmetry is a universal and fundamentally important property in nature, art and science (*17*). In physics, geometrical symmetry is associated with physical conservation, which is one of the most powerful tools of theoretical physics, as it has become evident that all laws of nature originate from symmetries (*18, 19*). Chirality is a type of asymmetry with the absence of reflection symmetry. It widely exists in nature ranging from chemical molecules (*20*) to biological organisms (*21*). Conventional ball-shaped geometries, such as spheres and ellipsoids, have both reflection and rotational symmetry (Fig. 1 (A)). As their two-dimensional (2-D) projections, the conventional circle and ellipse are nonchiral. However, some 2-D chiral patterns, for example, a shape in the form of the letter "*S*" (Fig. 1 (B)), possess not only chirality, but also elliptic rotational symmetry (*22*). Some metamaterials have been studied that exploit the symmetry of 2-D chiral patterns as artificial atoms (*23*). However, their three-dimensional (3-D) counterparts, namely "*chiral balls*", which possess both chirality and 3-D rotational symmetry along every axis of a Cartesian coordinate system (*S1*), have not previously been explored (Fig. 1 (C)).

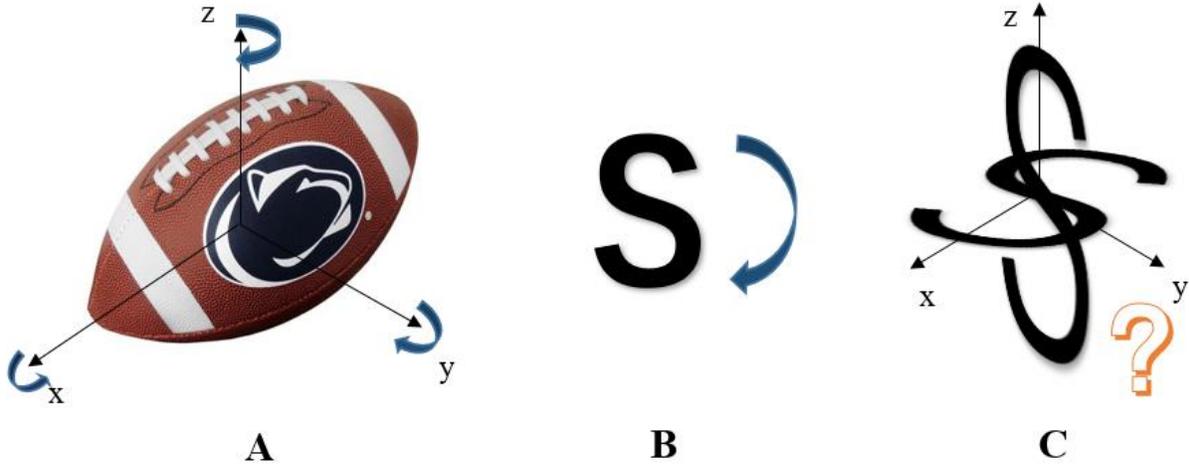

**Fig. 1**. Geometry of a conventional ball, a 2-D chiral pattern, and a 3-D extension of the 2-D chiral pattern. **(A)** A conventional ball has both reflection and rotational symmetry. **(B)** A 2-D chiral pattern shaped like the letter "*S*" has both chirality and rotational symmetry. **(C)** A simple 3-D extension of the "*S*" pattern does not have the expected rotational symmetry.

Here we observe that some specific torus knots possess such chiral ball symmetries. The definition of a knot is a closed curve in three-dimensional space that does not intersect itself anywhere. A conventional torus knot lies on the surface of a circular or elliptical torus in $\mathbf{R}^3$:

$$\begin{cases} x = (a + b \cdot \cos(q \cdot t)) \cdot \cos(p \cdot t) \\ y = (a + b \cdot \cos(q \cdot t)) \cdot \sin(p \cdot t) \\ z = c \cdot \sin(q \cdot t) \end{cases} \quad (1)$$

where $0 \leq t \leq 2\pi$. Parameter *a*, *b*, and *c* define the virtual torus (circular or elliptical) on which the knot wraps around. Integers *p* and *q* are positive (except for loops), and are relatively prime (*i.e.*, their greatest common divisor is one). Moreover, the integers *p* and *q* indicate the number of times the knot wraps around the torus in the longitudinal and meridional directions, respectively. A longitudinal curve runs the long way around the torus, while a meridional curve runs the short way (*14*). Any knot that can be continuously deformed into a circular loop

(standard ring) is said to be a trivial knot or unknot. Note that the torus knots are trivial if either $p$ or $q$ are equal to one.

All torus knots are chiral, as well as $q$-fold rotationally symmetric about the $z$-axis (*1*). To be rotationally symmetric also about the $x$- and $y$-axes, the parameter $q$ has to be even. Therefore, we reveal that the simplest chiral ball with the desired symmetry is a (3, 2)-torus knot. It is topologically equivalent to a (2, 3)-torus knot (*i.e.*, a trefoil), but geometrically distinct. Fig. 2 (A) depicts a perspective view of the knotted chiral ball. Fig. 2 (B), (C) and (D) are three views from the $x$, $y$, and $z$ axis, respectively. The knot is evidently chiral for its absence of reflection symmetry. Moreover, it is two-fold rotationally symmetric around every axis in a Cartesian coordinate system.

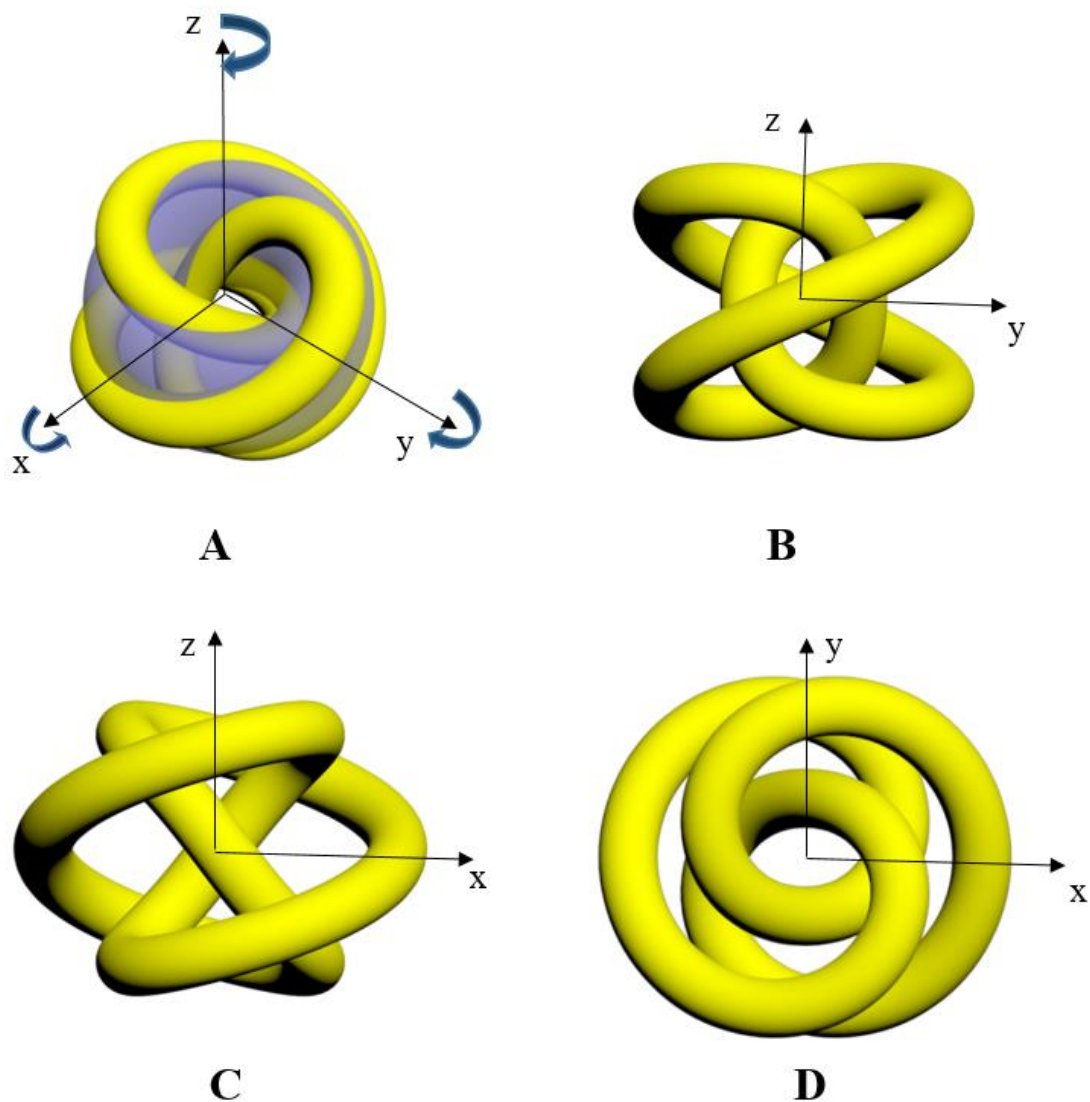

**Fig. 2**. Geometry of a nontrivial 3-D chiral ball. **(A)** Perspective view of a portion of a torus knot, functionalized as a 3-D chiral ball possessing both chirality and 3-D rotational symmetry. The knot (yellow) lives on the surface of a torus (purple). Three views of the knot from the **(B)** $x$,

**(C)** *y*, and **(D)** *z* axes. The nontrivial 3-D chiral ball is two-fold rotationally symmetric along the three axes of a Cartesian coordinate system.

Accordingly, Fig. 3 (A) shows the simplest trivial chiral ball as a (1, 2)-torus knot (*i.e.*, the unknot). Fig. 3 (B), (C) and (D) are three views from the *x*, *y*, and *z* axis, respectively. Although even without any inextricable knot, the trivial knot similarly possesses both chirality and 3-D rotational symmetry. This example also proves that the symmetry of chiral balls is independent of the knot topology.

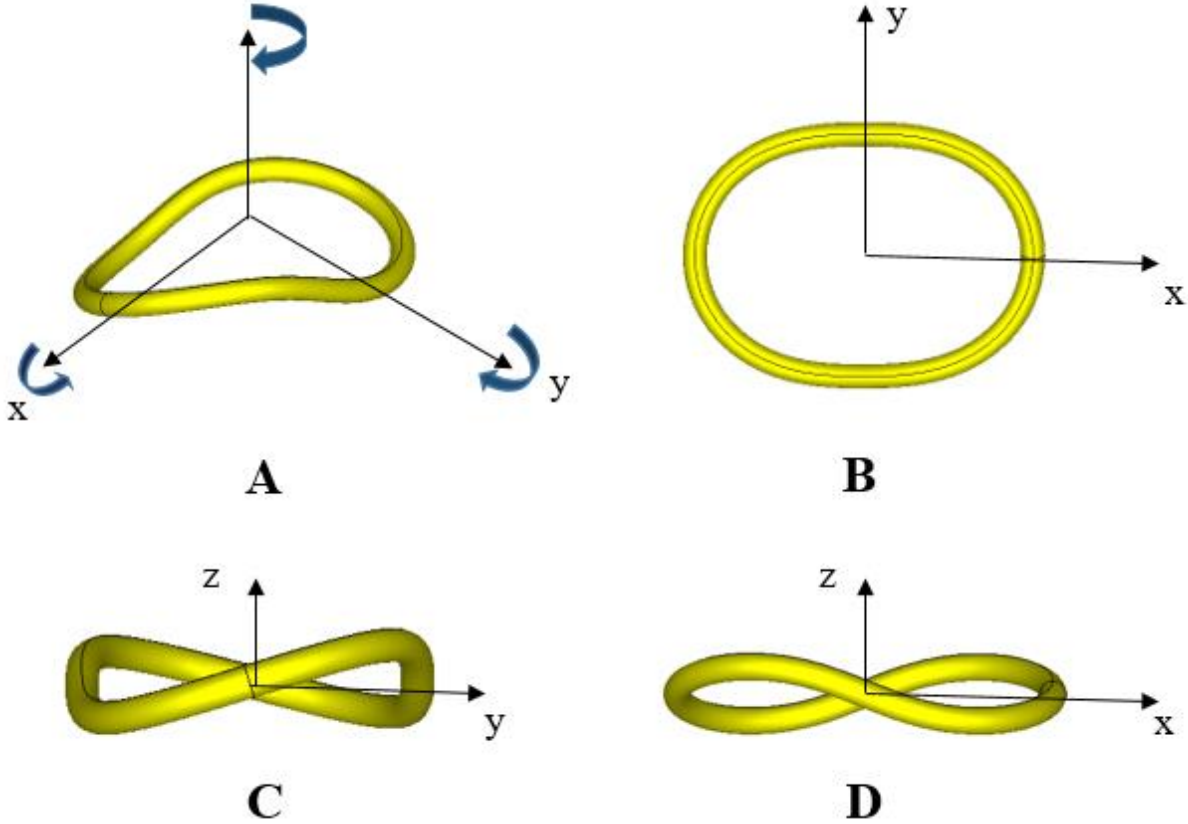

**Fig. 3**. Geometry of a trivial 3-D chiral ball. **(A)** Perspective view of a (1, 2)-torus knot, functionalized as a trivial 3-D chiral ball possessing both chirality and 3-D rotational symmetry. Three views of the knot from the **(B)** *x*, **(C)** *y*, and **(D)** *z* axes. The trivial 3-D chiral ball is two-fold rotationally symmetric along the three axes of a Cartesian coordinate system.

We also point out that distorting the chiral ball along the axes with different scaling factors will not break its chirality nor its rotational symmetry. The parameterization for the distorted torus knot can be expressed as:

$$\begin{cases} x = S_x \cdot (a + b \cdot \cos(q \cdot t)) \cdot \cos(p \cdot t) \\ y = S_y \cdot (a + b \cdot \cos(q \cdot t)) \cdot \sin(p \cdot t) \\ z = S_z \cdot c \cdot \sin(q \cdot t) \end{cases} \quad (2)$$

where $S_x$, $S_y$, and $S_z$ are the distortion scaling factors along the $x$, $y$, and $z$ axes, respectively.

As topologically stable objects, knots have attracted much attention in the fields of mathematics, physics, biology and engineering. In fact, in addition to their unique topological properties, this work reveals that knots also possess exotic symmetries that have not been previously investigated. This work suggests that further investigation into the symmetry of knots is merited, as well as its potential relationship with various fundamental physical phenomena ranging from elementary particle theory and molecular structure to cosmic textures found in the universe.

S1 is included in the Supplementary Material.

**Acknowledgments:** The authors would like to thank Danny Zhu, Ronald Jenkins, and Pingjuan Werner of Penn State University for their early support on fabrication of knot samples and discussions on potential methodologies for experimental validation. This work was supported in part by the Penn State MRSEC, Center for Nanoscale Science (NSF DMR-1420620), the National Excellent Youth Fund by NSFC (61425010), the Foundation for Innovative Research Groups of NSFC (61721001), the Changjiang Scholar Program, the EPSRC Program (EP/R013918/1), and the Sichuan Science and Technology Program (2018GZ0251).